\documentclass[twocolumn]{jpsj2} %% two-column layout

\title{%
Field Re-entrant Superconductivity Induced by the Enhancement of Effective Mass in URhGe%
}

\author{Atsushi \textsc{Miyake}$^{1,2}$\thanks{E-mail address: atsushi.miyake@cea.fr}, Dai \textsc{Aoki}$^{1}$, and Jacques \textsc{Flouquet}$^{1}$}

\inst{%
$^{1}$INAC/SPSMS, CEA-Grenoble, 17 rue des Martyrs, 38054 Genoble, France \\
$^{2}$KYOKUGEN, Osaka University, Toyonaka, Osaka 560-8531, Japan
}

\recdate{\today}

\abst{
High quality single crystals of a ferromagnetic superconductor URhGe were successfully grown.
The electrical resistivity was measured for the field along $b$-axis in the orthorhombic crystal structure
in order to study precisely its field re-entrant superconductivity which occurs in the vicinity of the field $H_{\rm R}$,
where the easy magnetization switches from $c$ to $b$-axis.
The field re-entrant superconductivity of URhGe is analyzed 
with special focus on its dependence with the value of residual resistivity $\rho_0$ 
and the coefficient $A$ of $T^2$ term of the resistivity.
The experimental results are well explained by a crude model 
related with the field dependence of the effective mass $m^\ast$,
where the corresponding critical temperature $T_{\mathrm{sc}}(m^\ast)$ 
and the upper critical field $H_{\rm c2}$ are strongly enhanced.
Discussion is made on the interplay between magnetic and superconducting phase diagram as well as the link between $T_{\mathrm{sc}}$ and $A$ in other heavy fermion superconductors.}
\kword{Field re-entrant superconductivity, URhGe, Ferromagnetic superconductor, Effective mass}

\begin{document}
\maketitle
\section{Introduction}
Ferromagnetism and superconductivity had been thought to be mutually competitive phenomena,
since the large internal field easily destroys the Cooper pair for conventional superconductivity.
In 1970s, the superconductivity was found in 4\textit{f}-localized ferromagnets,
such as ErRh$_4$B$_4$~\cite{Mon80,Fer77}, 
where the superconducting critical temperature $T_{\rm sc}$ is 
larger than the Curie temperature $T_{\rm Curie}$.
Below $T_{\rm Curie}$, the superconducting phase is expelled, thus the both phases 
are competing with each other.
In these systems with two separated electronic bath (localized 4\textit{f} electrons and light itinerant electrons),
the ferromagnetism is a robust quantity even for $T_{\rm Curie} < T_{\rm sc}$.
The discovery of superconductivity in the 5\textit{f}-itinerant ferromagnet UGe$_2$ under pressures was one of the breakthrough in the unconventional superconductivity \cite{saxena}.
Contrary to ErRh$_4$B$_4$, $T_{\rm sc}$ is lower than $T_{\rm Curie}$ in UGe$_2$
and the superconductivity exists only in the ferromagnetic phase on the pressure-temperature phase diagram,
where the superconductivity coexists with the ferromagnetism. 
The triplet superconductivity with equal spin pairing is, therefore, believed, 
because the large internal field will not prevent the formation of the Cooper pair.

Other examples for the coexistence of ferromagnetism and superconductivity are
URhGe~\cite{dai} and the recently found UCoGe~\cite{Huy,Has08}, which crystallize in the orthorhombic structure with TiNiSi-type. 
In the case of URhGe, the ferromagnetic (FM) transition occurs at $T_{\rm Curie}=9.5\,{\rm K}$ and its ordered moment
is $M_0=0.4\,\mu_{\rm B}/{\rm U}$.
The superconductivity (SC) was found to exist below $T_{\rm sc}=0.25\,{\rm K}$ at ambient pressure \cite{dai}.
The upper critical fields $H_{\rm c2}$ are larger than the Pauli limiting field,
thus the spin-triplet state is most likely realized \cite{hardy}.
Striking feature is the field re-entrant superconductivity (RSC)~\cite{Levy,Levy2}.
With increasing the magnetic field along the $b$-axis in the orthorhombic structure,
the moment starts to tilt from the easy magnetization $c$-axis to $b$-axis,
and then the ordered ferromagnetic moment is completely aligned to the $b$-axis
above a reorientation field $H_{\rm R}=12\,{\rm T}$.
The RSC is observed between $H_1=8$ and $H_2=12.7\,{\rm T}$ and detected in wide range of the field angle around $H \parallel b$.
Thus, this superconductivity is not due to the so-called Jaccarino-Peter effect \cite{Jaccarino-peter},
which has been found in the Chevrel phase compound~\cite{Meu84} and the organic superconductors, such as $\kappa$-(BETS)$_2$FeBr$_4$~\cite{Kon04},
where the RSC is due to a compensation of the external field by the internal exchange field.

An enhancement of $H_{\rm c2}$ related to the metamagnetic transition 
has been observed in the ferromagnetic superconductor UGe$_2$, as well.~\cite{Ilya}.
When the magnetic field is applied along the easy-magnetization axis ($a$-axis) 
at a pressure of $P=1.35\,{\rm GPa}$,
the ferromagnetic phase changes from FM1 (weakly polarized phase) to FM2 (strongly polarized phase) at $H_x$
with increasing field.
The $H_{\rm c2}$--$T$ phase diagram shows the step-like increase of $H_{\rm c2}$ at $H_x$ and
$H_{\rm c2}(0)$ in FM2 is higher than that expected in FM1.

% In URhGe, when the magnetic field increases along the $b$ axis 
% first the system transits from SC to a normal FM phase, and then 
% SC reappears at a magnetic field $H_1$ quite smaller than $H_{\mathrm{R}}$ \cite{Levy}.  
% Further increasing the fields, the SC is destroyed at $H_2\sim$12.7~T by extrapolating to 0~K. 
% Obviously, the field reorientation is associated with fluctuations 
% over a large field window $\Delta H_{\mathrm{RSC}}=H_2-H_1 \sim$4.7~T, which favors the restoration of SC.
%% It is revealed that $T_{\mathrm{SC}}$ of URhGe strongly depends on the quality of the sample, indicating the SC is not conventional but unconventional one. 

Here we report the careful studies on the field dependence of the inelastic $T^2$ term $A$
of the electrical resistivity $\rho$ by using the high quality single crystals
in order to study the RSC of URhGe.
The resistivity follows the quadratic temperature dependence, namely $\rho = \rho_0 + AT^2$ for all the measured field range.
The RSC is well explained by the enhancement of the effective mass $m^\ast$.

%which obeys whatever the magnetic field is a $T^2$ dependence, i.e. $\rho=\rho_0+AT^2$, where $\rho_0$ is the residual resistivity.
%The coefficient $A$ is known as that is proportional to the square of the effective electron mass, $m*$.

\section{Experimental}
The single crystals of URhGe were grown by the Czochralski pulling method in the radio frequency furnace 
under purified Ar atmosphere gas
with a stoichiometric ratio of the starting materials, U (Purity: 3N-$99.9\,{\%}$), Rh (4N) and Ge (6N).
The ingot was annealed at $1300\,^\circ {\rm C}$ under ultra high vacuum (UHV) for one day.
The single crystal ingot was oriented by the Laue photograph and was cut by the spark cutter.
The samples were annealed again at $900\,^\circ{\rm C}$ in the electrical furnace under UHV for 20 days.
The electrical resistivity was measured by the four probe AC method
at low temperatures down to $70\,{\rm mK}$ and at high magnetic fields up to $16\,{\rm T}$. 
The electrical current and magnetic fields were applied along $a$- and $b$-axis, respectively.
We carried out the resistivity measurements by using two single crystals.
The sample ({\#}1) with two pairs of voltage contacts ({\#}1a and {\#}1b) 
and the other sample with one pair of voltage contact ({\#}2)
are employed.
This allows us to compare two sets of data ({\#}1a and {\#}1b) with different residual resistivity in the same crystal without
misalignment between sample {\#}1a and sample {\#}1b.
The residual resistivity ratios (${\rm RRR} = \rho_{\rm RT}^{\mathstrut} / \rho_0$) are 40, 16 and 12, respectively, 
for sample {\#}1a, {\#}1b and {\#}2, indicating the high quality single crystals.

\section{Results and Discussion}
%---------------------------------------------------------------
\begin{figure}[tb]
\begin{center}
\includegraphics[width=0.9 \hsize,clip]{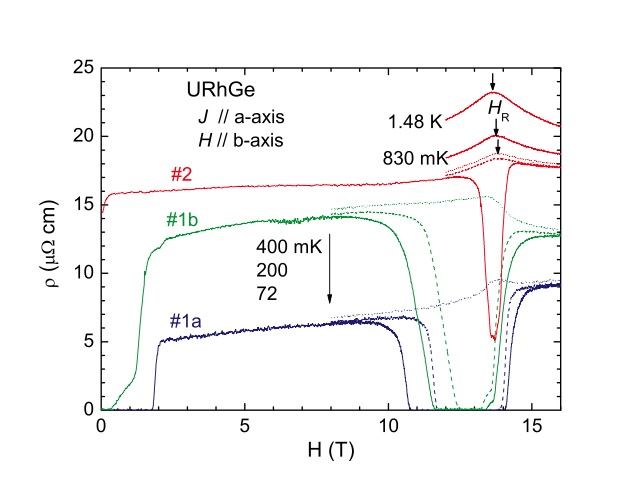}
\end{center}
\caption{
(Color online)
Field dependences of the resistivity at constant temperatures in sample {\#}1a, {\#}1b and {\#2} for the current along $a$-axis and the field along $b$-axis.
The solid, dashed and dotted lines show the data at 72, 200 and $400\,{\rm mK}$, respectively.
For sample {\#}2, the data at 830\,mK and 1.48\,K are displayed as well.
The arrows indicate the reorientation field $H_{\rm R}$.
}
\label{f1}
\end{figure}
%---------------------------------------------------------------
Figure~\ref{f1} represents the field dependence of the resistivity $\rho$ for {\#}1a, {\#}1b and {\#}2 at 72, 200 and $400\,{\rm mK}$.
For sample {\#}1a with ${\rm RRR}=40$, the RSC was clearly observed 
between $H_1=10.7\,{\rm T}$ and $H_2=14.2\,{\rm T}$ at $72\,{\rm mK}$,
where $H_1$ and $H_2$ are defined as mid points of the drop of resistivity.
At high temperature of $400\,{\rm mK}$, the resistivity shows the peak at $H_{\rm R}=13.7\,{\rm T}$,
corresponding to the reorientation of the moment.
The crystal seems to misalign slightly from the field direction respect to the $b$-axis, 
because $H_1$, $H_2$ and $H_{\rm R}$ are larger than those in the previous reports~\cite{Levy,Levy2}.
The purer sample, i.e. higher value of RRR, shows the wider superconducting windows and the higher transition temperatures.
It is interesting to note that for sample {\#}2, only track of SC is suspected at low field, while rather deeper drop on the resistivity is observed in a very narrow field window.
The same trend is observed in sample {\#}1b,
where the resistivity shows rather sharp drop due to the RSC,
although the step-like behavior is observed near $H_{\rm c2}$ for the low field SC.
These observation indicates the strict condition for RSC.
That suggests a strong increase of the coherence length in the RSC phase.
On the other hand, the reorientation field $H_{\rm R}$ are not sensitive to the sample quality.
The insensitivity is also a FM property at zero field \cite{dai}.
It is noted that $H_1$ is more sensitive to the sample quality compared to $H_2$,
indicating that the RSC strongly stick to $H_{\rm R}$.

\begin{figure}[tb]
\begin{center}
\includegraphics[width=0.9 \hsize,clip]{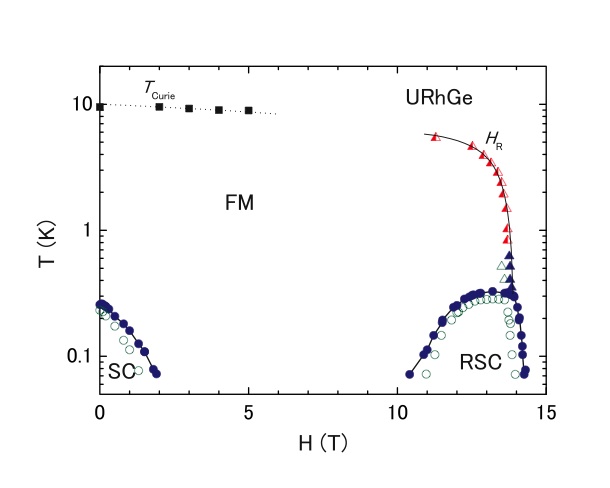}
\end{center}
\caption{
(Color online) Temperature-field phase diagram of the low field superconductivity (SC), the re-entrant SC (RSC) and reorientation field $H_{\rm R}$ for $H \parallel b$ in URhGe.
The circles display the critical field (temperature) of the superconductivities defined as a midpoint of the resistivity drop. 
The triangles correspond to the reorientation field $H_{\rm R}$ defined as a peak of the $\rho(H)$ curves. 
The solid squares are ferromagnetic transition temperature $T_{\rm{Curie}}$ cited from Ref.~\protect\citen{Hux}.
The dark blue (filled symbol), green (open) and red (half filled) correspond to the results for sample {\#}1a, {\#}1b and {\#}2, respectively.
The lines are guides for eyes.}
\label{fig:HT_phase}
\end{figure}

\begin{figure}[tb]
\begin{center}
\includegraphics[width=0.9 \hsize,clip]{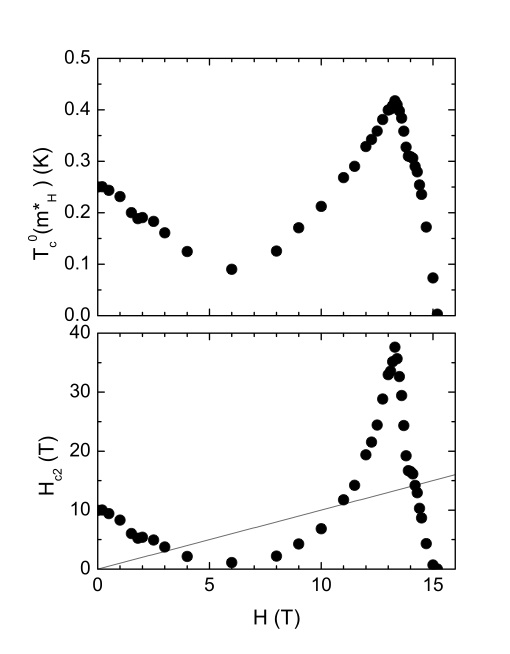}
\end{center}
\caption{
(a)Calculated $T_{\rm sc}$ at zero field and (b)$H_{\rm c2}$ as a function of magnetic field. 
The solid line in panel (b) indicates the applied magnetic field. 
The superconductivity appears under the condition of $H_{\rm c2} > H$.
}
\label{f5}
\end{figure}

Figure~\ref{fig:HT_phase} shows the superconducting phase diagram and reorientation field $H_{\rm R}$.
In sample {\#}1a, the upper critical field $H_{\rm c2}(0)=2\,{\rm T}$
is clearly larger than the Pauli limiting field $H_{\rm p}=0.46\,{\rm T}$ estimated from $T_{\rm sc}(0)=0.25\,{\rm K}$,
indicating that $H_{\rm c2}$ is determined by the orbital limit depending on the effective mass.
The RSC is observed between $H_1=10.7$ and $H_2=14.2\,{\rm T}$ at $72\,{\rm mK}$ for sample {\#}1a.
It is interesting that
$T_{\rm sc} (H)$ for RSC reaches a temperature of $330\,{\rm mK}$ at $13.3\,{\rm T}$,
which is larger than the value of $T_{\rm sc}(0)=250\,{\rm mK}$ at zero field.
This peculiar behavior is consistent with the analysis of $T_{\rm sc}$ by the
enhancement of effective mass $m^\ast$ in Fig.~\ref{f5} as discussed later.
At high temperatures, the reorientation field $H_{\rm R}$ starts to decrease with increasing temperature.
Finally, $H_{\rm R}$ seems to be connected to the Curie temperature $T_{\rm Curie}=9.5\,{\rm K}$ at zero field.
%---------------------------------------------------------------
%---------------------------------------------------------------
%---------------------------------------------------------------
\begin{figure}[ht]
\begin{center}
\includegraphics[width=0.9 \hsize,clip]{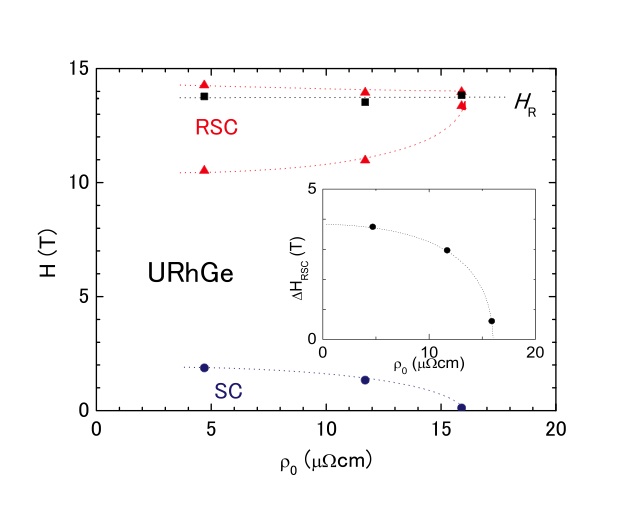}
\caption{
(Color online)
Critical fields of superconductivities, $H_{\mathrm{c2}}$, $H_{1}$ and $H_2$, and the reorientation field $H_{\rm R}$ as a function of the residual resistivity $\rho_0(H = 0)$ at 72mK in URhGe. The residual resistivity is inversely proportional
to the mean free path $l$.
The dark blue circle and red triangle are the critical fields of low field and re-entrant
superconductivities, respectively. 
The black square is reorientation field $H_{\rm R}$. 
The inset shows the field width of RSC, $\Delta H_{\rm RSC}$. 
The dotted lines are guides for eyes.}
\label{f2}
\end{center}
\end{figure}
%-------------------------------------------------------------------

Figure \ref{f2} represents the variations of $H_{\mathrm{c2}}$, $H_{1}$, $H_2$ and $H_{\rm R}$ at 72~mK as a function of $\rho_0$, which is inversely
proportional to the mean free path, i.e. $\rho_0 \propto 1/l$. For unconventional superconductivity like URhGe, the
$T_{sc}$ must depend on the variation of the parameter $\xi/l$ between the superconducting coherence length $\xi$
and $l$ according to the Abrikosov-Gor'kov pair breaking mechanism~\cite{abrikosov}, where $l$ must be larger than the
coherence length $\xi$. The so-called clean limit condition must be satisfied. Such a dependence has been actually observed in previous studies on polycrystalline samples~\cite{Aok03} and single crystals~\cite{hardy}. 
The striking point is that the RSC also collapses when the low field SC collapses, moreover, the RSC width $\Delta H_{\rm{RSC}}$ strongly depends on $\rho_0$.
The suppression of RSC with lower quality samples indicates that 
RSC as well as low field SC are unconventional.

% As shown in Fig.~\ref{f3},
% these data are in good agreement with the measured field variation of $\rho_0 (\propto 1/{\rm RRR} \propto 1/l)$ 
% and of $A (H)$, which is proportional to the square of effective mass $m^\ast (H)$.
% The ratio $\xi/l$ must be weakly pressure dependent.

\begin{figure}[tb]
\begin{center}
\includegraphics[width=0.9 \hsize,clip]{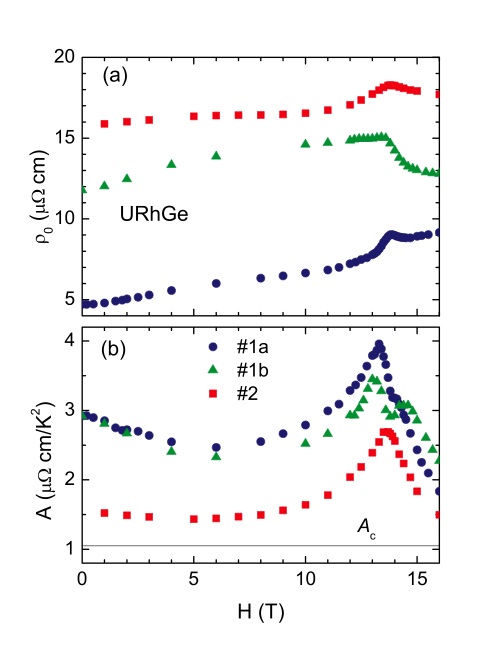}
\end{center}
\caption{
(Color online)
Field dependence of (a) the residual resistivity $\rho_0$ and (b) the coefficient $A$ on URhGe for sample \# 1a, \# 1b and \# 2.
The values are obtained by the least square fit with $\rho (T) = \rho_0 +AT^2$ between $T_{\rm sc}$ and $750\,{\rm mK}$. 
The horizontal line denoted by $A_{\rm{c}}$ in panel (b) corresponds to the square of the band mass, ($m_{\rm B}^{\mathstrut}{})^{2}$, using the calculation of $T_{\mathrm{sc}}^0(m^*_H)$ and the detail is in the text.}
\label{f3}
\end{figure}

Here we simply estimate $T_{\rm sc}$ for RSC from strength of the parameter $\xi/ l$.
Since the coherence length $\xi$ is given by the relation, $\xi \sim \hbar v_{\rm F}^{\mathstrut} / k_{\rm B} T_{\rm sc}$,
where $v_{\rm F}^{\mathstrut}$ is Fermi velocity, 
$\xi$ is inversely proportional to $m^\ast T_{\mathrm{sc}}$.
If the $\xi/l$ is assumed to be invariant against the magnetic field,
$\xi/l \propto 1/(m^\ast l T_{\rm sc}) \propto \rho_0 / (m^\ast T_{\rm sc}) \propto \rho_0/(\sqrt{A}T_{\rm sc})$ 
must be a constant,
we can estimate the extrapolated value at zero field of the critical temperature $T^0_{{\rm sc}} (m^\ast_{H_{\rm R}})$ for fictitious quasiparticles of mass $m^\ast_{H_{\rm R}}$ at $H = 0$.
% we can estimate $T^0_{{\rm sc}} (m^\ast_{H_{\rm R}})$ for RSC, which is the value extended to the zero field.
Here we assume the Kadowaki-Woods relation between $m^\ast$ and $A$, namely $m^\ast \sim \sqrt{A}$.
From Fig.~\ref{f3}, one can obtain the ratio $(\rho_0 / \sqrt{A})_{H=H_{\rm{R}}} /(\rho_0 / \sqrt{A})_{H=0} \sim 1.4$
%$(m^\ast l)_{H=0}/(m^\ast l)_{H=H_{\rm R}} \sim 1.4$,
and then $T^0_{\rm sc} (m_{H_{\rm R}}^\ast) \sim 1.4 T^0_{\rm sc} (m_{H=0}^\ast)$:
$T^0_{\rm sc} (m_{H_{\rm R}}^\ast)$ can be estimated as $350\,{\rm mK}$ from $T^0_{\rm{sc}}(m^\ast_{H=0})=250\,{\rm mK}$.
Consistent with this zero field extrapolation, the experimental result in Fig.~\ref{fig:HT_phase}
indicates $T_{\rm sc}(m_{H_{\rm R}}^\ast, H_{\rm R}) \simeq 320\,{\rm mK}$ for sample {\#}1a,
which is smaller than the evaluated $T_{\rm sc}(m_{H_{\rm R}}^\ast, H=0)$.

%
%If the reference of $T_{\rm sc}(m^\ast)$ at zero field will not change, $\xi/l$ will vary on $\rho_0/\sqrt{A}$.
%The ratio changes from $H=0$, to $H_{\mathrm{R}}$ by only a factor of 1.4 for sample 1a: i.e. $(\rho_0/\sqrt{A})_{H_{\mathrm{R}}}/(\rho_0/\sqrt{A})_{H=0} \sim 1.4$.
%We see the factor that the clean observation of re-entrant SC requires to take into account an enhancement by at least a factor 1.4 of $T_{c}(m^*_{H=0})$ (Eq. \ref{eq2}), which is the calculated $T_{\mathrm{SC}}$ at zero field and discusses later.
%In agreement with published result, 

%---------------------------------------------------------------
%---------------------------------------------------------------
%---------------------------------------------------------------
\begin{figure}[ht]
\begin{center}
\includegraphics*[width=0.9 \hsize,clip]{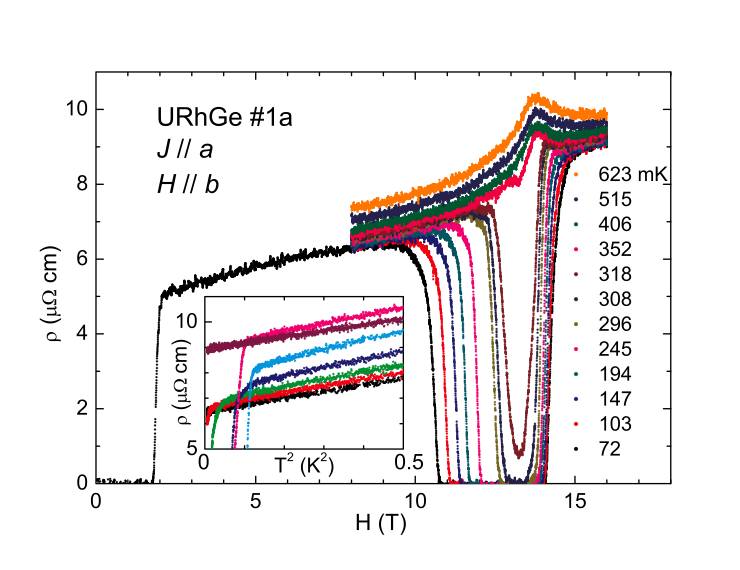}
\caption{(Color online)Field dependence of the resistivity on sample {\#}1a at various temperatures. The numbers of legend indicate temperatures in the unit of mK.
The inset shows $T^2$-dependence of the resistivity under magnetic fields of $H$~=9, 10, 11, 12, 13, 15 and 14~T from bottom to top.}
\label{f4}
\end{center}
\end{figure}
%---------------------------------------------------------------

As shown in Fig.~\ref{f3}(b), the $A$ coefficient starts to decrease slightly and increase again with increasing field.
It has a maximum just below $H_{\rm R}$, where $\rho_0$ has a maximum (Fig.~\ref{f3}(a)), then strongly decreases.
% At first glance, there \blue{may have} a critical value of $A=A_{\rm c}$ for superconductivity.
The effective mass may play an important role for the appearance of superconductivity.
Thus we need to evaluate what will be the SC temperature $T^0_{\rm sc}(m^\ast_H)$,
assuming that the effective mass is no more the effective mass at zero field, $m_{H=0}^\ast$,
but the mass $m_H^\ast$ built by the $H$ dressing of the quasiparticle.

It is well known that the calculation of $T_{\rm sc} (m^\ast)$ is a difficult task
which requires the solution of the Eliashberg equation.
However, we will show that the RSC in URhGe seems to be well explained by a crude model
where $T_{\rm sc}$ is related to the field variation of the effective mass $m^\ast_H$.
The simplicity of URhGe comes from its non-proximity to the ferromagnetic instability;
the pressure ($P$) increases $T_{\rm Curie}$, and decreases $T_{\rm sc}$ as well as the $A$ coefficient.~\cite{P-T_phased, thesis_FH}
$P$ and $H$ scans show that URhGe corresponds to the simple case where the variation of the $A$ coefficient 
(basically $m^\ast$) and $T_{\rm sc}$ are coupled, that is,
the enhancement of $A$ leads to an enhancement of $T_{\rm sc}$.
The chosen expression of $T_{\rm sc}^0 (m^\ast_H)$ is 
\begin{equation}
T_{\rm sc}^0 (m^\ast_H) = T_0 \exp\left( -\frac{m^\ast_H}{g m^{\ast\ast}} \right),
\end{equation}
where the effective mass $m^\ast_H$ is described by the band mass $m_{\rm B}^{\mathstrut}$
plus an extra mass $m^{\ast\ast}$ directly related to the source of SC pairing, namely,
\begin{equation}
m^\ast_H = m_{\rm B}^{\mathstrut} + m^{\ast\ast}.
\end{equation}
Assuming $g=1$ for simplicity,
\begin{align}
T_{\rm sc} (m_H^\ast) &= T_0 \exp\left[ -\left(\frac{m_{\rm B}^{\mathstrut}}{m^{\ast\ast}} + 1\right)\right] 
\label{eq1}
\end{align}
The expression is based on the McMillan-like formula~\cite{McMillan}; 
for FM superconductors, it was theoretically proposed far from FM instability.~\cite{Fay,Kir01,Sandeman}
Taking the logarithmic derivative of eq.\,(\ref{eq1}), it is interesting to remark that
Gr\"{u}neisen parameter for $T_{\rm sc}$ will be linked to the Gr\"{u}neisen parameter of
the effective temperature $T_{\rm B} \sim 1/m_{\rm B}^{\mathstrut}$ and $T^{\ast\ast} \sim 1/m^{\ast\ast}$ as,
\begin{equation}
\Omega_{T_{\rm sc}} = \frac{1}{\lambda} (\Omega_{T_{\rm B}} - \Omega_{T^{\ast\ast}} ),
\end{equation}
where Gr\"{u}neisen parameter is defined as $\Omega_T = - d(\log T)/d(\log V)$, and $\lambda = m^{\ast\ast}/m_{\rm B}^{\mathstrut}$. 
If the main phenomena is due to the volume dependence of $T_{\rm B}$, that is,
the systems are mainly dependent on the $\Omega_{T_{\rm B}}$,
$T_{\rm sc}$ and $T_{\rm B}$ will go to the same direction.
In other word, $T_{\rm sc}$ will increase when the band mass $m_{\rm B}^{\mathstrut}$ decreases.
Basically it is the main argument that considering different strongly correlated electron system 
from heavy fermion systems to high $T_{\rm sc}$ oxides,
$T_{\rm sc}$ scales the inverse of the specific heat coefficient $\gamma$~\cite{Uemura,Moriya};
basically SC can appear only below the Fermi temperature which is related to $\gamma^{-1}$.
If the main phenomena is the volume dependence of $T^{\ast\ast}$, which is observed
in uranium heavy fermion compounds like UPt$_3$, URu$_2$Si$_2$ and UBe$_{13}$,
the Gr\"{u}neisen parameter of the SC and normal phases have opposite sign \cite{JFreview}. 
The same trend is observed in URhGe.

Assuming the relation $A\sim (m^\ast)^2$,
and that the upper critical field $H_{\rm c2}(H) _{T\to 0}$ is governed by the orbital limit for this triplet 
superconductor, $H_{\rm c2}(H) _{T\to 0}$ is given by the equation,
\begin{equation}
H_{\rm c2} (H)_{T\to 0} \sim  \left[ m^\ast (H) T_{\rm sc}^0 (m^\ast_H) \right]^2,
\end{equation}
one can estimate $T_{\rm sc}^0(m^\ast_H)$ and $H_{\rm c2}(H)$ for a given value of $A_{\rm c} \sim (m_{\rm B}^{\mathstrut})^2$.
Figure~\ref{f5} represents the predicted $H$ dependence
of $T_{\rm sc}^0(m^\ast_H)$ and $H_{\rm c2}(m^\ast_H)$ as a function of $H$ for sample {\#}1a,
where $A_{\rm c}=1.1\,\mu\Omega\, {\rm cm}/{\rm K^2}$ is chosen, 
corresponding to the hypothesis of a field independent band mass $m_{\rm B}^{\mathstrut}$.
Following our suggestions~\cite{JF_private}, similar calculations were made~\cite{{thesis_FL}}, however, 
no measurements were carried out for the field dependence of $T^2$-law in the resistivity.
As shown in the inset of Fig.\ref{f4}, the quadratic temperature dependence of the resistivity were observed at high fields.
%For all the measured fields, no deviation from the Fermi-liquid law of $T^2$ dependence was observed. 

Surprisingly, the predicted $H_{c2}$ is in good agreement with the results of experiments, as shown in Fig.~\ref{f5}(b).
$T_{\mathrm{sc}}$ at $H_{\mathrm{R}}$ is enhanced by a factor of 1.7, 
which is larger than the estimated value from the field variation of $\rho_0/\sqrt{A}$, as discussed above.
Here we note that the $T_{\rm sc}^0(m^*(H))$ strongly depends on the chosen value of $A_{\rm c}$, namely $m_{\rm B}^{\mathstrut}$. 
When $m_{\rm B}^{\mathstrut}$ increases, the evaluated $T_{\mathrm{sc}}^0(m^*(H))$ increases as well.

%---------------------------------------------------------------
%---------------------------------------------------------------

%---------------------------------------------------------------
\begin{figure}[tb]
\begin{center}
\includegraphics[width=0.9 \hsize,clip]{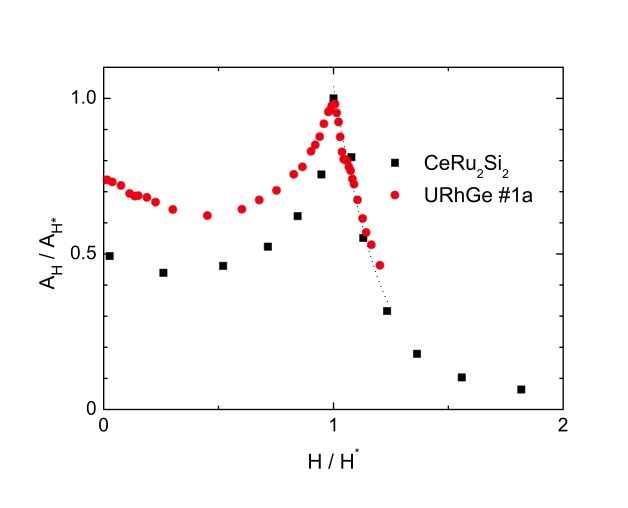}
\end{center}
\caption{(Color online) Relative field variation of $A_H/A_{H^\ast}$ of URhGe (\#1a) and CeRu$_2$Si$_2$ as a function of a normalized field $H/H^\ast$. The data of CeRu$_2$Si$_2$ is cited from Ref.\,\citen{CeRh2Si2}. }
\label{f6}
\end{figure}
%---------------------------------------------------------------

URhGe is a good example where the RSC due to the spin reorientation can be
described by a field enhancement of the correlation, namely enhancement of $m^{\ast\ast}$ and invariance of $m_{\rm B}^{\mathstrut}$.
The spin reorientation mechanism leads to escape from the rigidity of Ising spin dynamics.
The fluctuation between $c$ and $b$-axis component of the magnetization
in the large field range may lead to a large field window of the mass enhancement,
which is in agreement with the large field region of RSC.
According to the neutron scattering experiments for $H\parallel b$, the sublattice magnetization for $c$-axis, $M_c$
starts to decrease from $8\,{\rm T}$ and becomes zero at $H_{\rm R}=11.5\,{\rm T}$ with increasing field.~\cite{Levy}
On the other hand, the sublattice magnetization for $b$-axis, $M_b$ linearly increases from zero,
and becomes $M_b = M_c$ at $10.5\,{\rm T}$. 
Finally, $M_b$ is fully polarized at $H_{\rm R}=11.5\,{\rm T}$.
It means that the fluctuation between $M_b$ and $M_c$ starts to develop from $8\,{\rm T}$.
Simply thinking, when $M_b = M_c$ is realized, the magnetic fluctuation may become maximum,
implying the large enhancement of the effective mass. 
This may correspond to a maximum value of $A$ at a field, which is slightly lower than $H_{\rm R}$ defined as a peak of $\rho_0$,
as shown in Fig.~\ref{fig:HT_phase}.
A fit of the resistivity $\rho$ by a $T^2$ law lead to find another weak maxima of $A$ for sample {\#}1a and {\#}1b.
Its origin is left to the future study. 
It may be caused by the intrinsic effect due to a $H$ crossover between collision and quantum criticality regime or the small misalignment inside the crystal.

On the other hand, for the other FM superconductor UGe$_2$, the field change of SC properties corresponds to a drastic switch in the 
description of the ferromagnetism: from perfectly polarized FM2 to imperfectly polarized FM1 according to the recent neutron scattering analysis \cite{aso}.
For UGe$_2$, the Ising character is preserved at the switch from FM1 to FM2 at $H_x$, 
furthermore this switch is associated with a drop of $m^\ast$.~\cite{tateiwa2001,Phi08}
The key phenomena for SC does not appear in the proximity to the ferromagnetic-paramagnetic (FM-PM) instability
but in the instability at $P_x$,
where the system may switch from FM2 to FM1.
The pressure dependence of $T_{\rm sc}$ cannot be related to the pressure increase of $\gamma$ term from FM2 to FM1,
via an increase of $m^{\ast\ast}$ and application of the eq.~(\ref{eq1}).
Obviously, the main phenomena is either a change in the Fermi surface topology as discussed in ref.~\citen{FS}
or another origin of pairing as proposed in the charge density wave scenario~\cite{Watanabe,Watanabe2}.
At least in the frame of our model, $T_{\rm B}$ must change on both side of $P_x$.
It is more complicated than the case of URhGe.
The drastic change of Fermi surfaces in UGe$_2$ were reported between $P < P_x$ and $P > P_{\rm c}$,
but the experimental results between $P_x$ and $P_{\rm c}$ are still unclear.~\cite{Terashima,Settai}

%=========
It is interesting to compare URhGe with CeRu$_2$Si$_2$, a heavy fermion compound
highly studied for its metamagnetic transition at $H_{\rm M}$
where a sharp crossover occurs between a low field paramagnetic phase (PM) and a high field polarized paramagnetic (PPM) phase
at $H_{\rm M}=7.7\,{\rm T}$.
Figure~\ref{f6} shows the relative field variation of $A_H/A_{H^\ast}$ in reduced scale of $H/H^\ast$,
where $H^\ast$ corresponds to $H_{\rm R}$ and $H_{\rm M}$ for URhGe and CeRu$_2$Si$_2$, respectively.
A perfect scaling occurs for $H/H^\ast > 1$, here both cases are governed by Ising ferromagnetic spin dynamics.
Below $H/H^\ast < 1$, the systems are quite different. However, enhancement of $A(H^\ast)$ are 
rather similar, $A(H^\ast)/A(0)=1.4$ and 2 for URhGe and CeRu$_2$Si$_2$, respectively.

Below $H_{\rm M}$, microscopic phenomena are now well clarified in CeRu$_2$Si$_2$~\cite{CeRu2Si2}.
Antiferromagnetic (AF) correlation dominates below $H_{\rm M}$ and collapses just at $H_{\rm M}$,
where the FM fluctuation becomes soft in a narrow field window centered at $H_{\rm M}$.
As discussed in ref.~\citen{HFroad}, an interesting situation would occur,
if CeRu$_2$Si$_2$ were SC at $H=0$ with $H_{\rm c2}^\ast > H_{\rm M}$.
Assuming the switch with the fluctuation from AF to FM, the singlet pairing may be replaced by the triplet pairing. 
Unfortunately, the Ising character of the AF spin dynamics presumably prevents the establishment of SC.

Thus the simplicity of URhGe is that the FM mechanism is preserved; furthermore
as $T_{\rm Curie}>T_{\rm sc}$, the $P$ and $H$ dependence of an unique parameter $m^{\ast\ast}$ appears
sufficient to describe the SC properties.
Up to now for AF heavy fermion systems, only $H$ reentrance of AF in the SC state of CeRhIn$_5$
has been reported \cite{Park, Knebel}; it corresponds to hierarchy between the bare parameters
($T_{\rm N}$: N$\acute{\rm{e}}$el temperature, $T_{\rm sc}$, $H_{\rm c}$: critical magnetic field of the AF--PM boundary and $H_{\rm c2}$):
$T_{\rm sc} > T_{\rm N}$ and 
$H_{\rm c} > H_{\rm c2}(0)$;
AF survives the normal phase up to $H_{\rm c}$.

An interesting case will be that $H_{\rm c}$ is lower than $H_{\rm c2}$, when $T_{\rm sc} > T_{\rm N}$.
That may happen for CeCoIn$_5$ as no AF is detected above $H_{\rm c2}(0)$,
however, a new low temperature-high field superconducting phase is observed.
In the framework of a magnetic scenario, one can imagine that the persistence of a SC 
gap inhibits the transition to PM phase which requires the collapse of the AF pseudogap; due to the interplay with SC,
AF may be sticked to $H_{\rm c2}(0)$.
Up to now, the main trend is 
to neglect the possible magnetic origin, and
to consider that the new phase of CeCoIn$_5$ is the evidence of the so-called FFLO state predicted
four decades ago with a extra modulation of the order parameter along $H$~\cite{Matsuda,FF,LO}.
But still no definitive conclusion emerges from the experiments.~\cite{You07}

\section{Conclusion}
The reorientation of the moment from $c$ to $b$-axis for $H \parallel b$ is 
characterized by an invariance of $H_{\rm R}$ on sample purity.
While the RSC seems to strongly depend as its low field SC on the sample purity,
indicating the unconventional nature for both RSC and low field SC.
The $T^2$ Fermi liquid law of the resistivity 
is obeyed for all the field range, the amplitude of the $A$ coefficient has an enhancement at $H_{\rm R}$
which is interpreted as an enhancement of the effective mass $m^\ast$.
The enhancement of $m^\ast$ appears to concern mainly the effective mass contribution $m^{\ast\ast}$
added by the magnetic correlation on the band mass $m_{\rm B}^{\mathstrut}$.
The increase of $m^{\ast\ast}$ in the vicinity of $H_{\rm R}$ leads to a strong increase of 
the superconducting transition temperature $T_{\rm sc}$ and of the upper critical field $H_{\rm c2}\sim (m^\ast T_{\rm sc})^2$.
Surprisingly, a crude model assuming McMillan-type formula for $T_{\rm sc}$ gives a good description 
of the experimental results.
It is interesting to mention that the enhancement of the fluctuation is quite non symmetrical with respect to $H_{\rm R}$.
The feedback on SC is the asymmetry of RSC boundary by reference to $H_{\rm R}$.
This property is clearly demonstrated in Fig.~\ref{f6} by the comparison with CeRu$_2$Si$_2$.
The difference of re-entrant phenomena between in URhGe and in UGe$_2$
has been discussed. The particularity of UGe$_2$ is to involve SC pairing
between quite different FM phases (fully polarized and partially polarized).
The main phenomena appear to be the change in Fermi surface topology.
At least for both cases, only FM is considered.
For the re-entrant phenomena on appearance of a new low temperature and high field phase as observed in so-called Ce115 compounds,
the antiferromagnetic fluctuations are dominant at $H=0$.
However, a field sweep will reveal the interplay between AF, PM, PPM and SC phase.
CeRhIn$_5$ appears a simple case where the field re-entrance of AF disappears
since the magnetic critical field $H_{\rm c}$ collapses rapidly under pressure.
CeCoIn$_5$ is a more intriguing example,
where AF may be sticked to $H_{\rm c2}(0)$ over a rather large $P$ window;
as indicated an alternative explanation is the formation of a FFLO state.

\section*{Acknowledgements}
We thank G. Knebel, J. P. Brison and F. Hardy for experimental supports and valuable discussion. 
A. M. has been supported by Research Fellowship of the Japan Society for the Promotion of Science for Young Scientists and EGIDE.
This work was financially supported by French ANR projects ECCE and CORMAT.

\end{document}